\documentclass[sigconf,screen,nonacm=true]{acmart}
\AtBeginDocument{%
  }

\usepackage{xspace}
\xspaceaddexceptions{\%}
\xspaceremoveexception{-}

\usepackage{colortbl}
\usepackage{xcolor}
\definecolor{gray}{rgb}{0.5,0.5,0.5}
\definecolor{mauve}{rgb}{0.57, 0.37, 0.43}
\definecolor{rltred}{rgb}{0.5,0,0}
\definecolor{rltgreen}{rgb}{0,0.5,0}
\definecolor{rltblue}{rgb}{0,0,0.5}
\definecolor{webgreen}{rgb}{0, 0.5, 0} 
\definecolor{webblue}{rgb}{0, 0, 0.5} 
\definecolor{webred}{rgb}{0.5, 0, 0} 
\definecolor{light-gray}{HTML}{B3B3B3}
\definecolor{dim-gray}{HTML}{696969}
\definecolor{white-close}{HTML}{E2E2E2}
\definecolor{JIAWEIdraft}{rgb}{0.24, 0.6, 0.45}
\definecolor{light-red}{rgb}{0.968627451,0.8078431373,0.8039215686} 
\definecolor{light-green}{rgb}{0.8392156863,0.9921568627,0.8156862745} 
\definecolor{light-gray}{rgb}{0.8745098039,0.8745098039,0.8745098039} 
\definecolor{defcolor}{HTML}{2B7A78} 
\usepackage[most]{tcolorbox}
\usepackage{wrapfig}

\usepackage{graphicx}
\usepackage{subcaption}
\usepackage{booktabs}
\usepackage{multirow}
\usepackage{tabularx}
\usepackage[hang,flushmargin]{footmisc} 

\usepackage{amsfonts}
\usepackage{amsmath}
\usepackage{amsthm}

\usepackage{enumitem}
\setlist[enumerate]{leftmargin=*}
\setlist[itemize]{leftmargin=*}

\usepackage{url}
\usepackage[capitalise,noabbrev]{cleveref} 
\crefname{lstlisting}{listing}{listings}
\Crefname{lstlisting}{Listing}{Listings}

\usepackage{listings}%
\usepackage{soul} 
\DeclareRobustCommand{\lightredhl}[1]{{\sethlcolor{light-red}\hl{#1}}}
\soulregister{\lightredhl}{1}
\DeclareRobustCommand{\lightgreenhl}[1]{{\sethlcolor{light-green}\hl{#1}}}
\soulregister{\lightgreenhl}{1}
\DeclareRobustCommand{\lightgrayhl}[1]{{\sethlcolor{light-gray}\hl{#1}}}
\soulregister{\lightgrayhl}{1}

\newcommand{\lstbg}[2]{\fboxsep=0pt\colorbox{#1}{\strut{#2}}}
\lstdefinelanguage{diff}{
  morecomment=[f][\color{lightgray}]{@@},
  morecomment=[f][\lstbg{light-green}]{+\ },
  morecomment=[f][\lstbg{light-red}]{-\ },
}

\usepackage[framemethod=TikZ]{mdframed}
\newcounter{lem}[section]
\renewcommand{\thelem}{\arabic{lem}}

\usepackage[braket,qm]{qcircuit}

\newcommand{\ourTool}{\textsc{QaCoCo}\xspace}
\newcommand{\ourToolFull}{Quantum Controlled Gate Coverage (\ourTool)\xspace}
\newcommand{\ourToolURL}{\url{https://github.com/QBugs/QaCoCo}\xspace}

\setcopyright{acmlicensed}
\copyrightyear{2018}
\acmYear{2018}
\acmDOI{XXXXXXX.XXXXXXX}
\acmConference[Conference acronym 'XX]{Make sure to enter the correct
  conference title from your rights confirmation email}{June 03--05,
  2018}{Woodstock, NY}
\acmISBN{978-1-4503-XXXX-X/2018/06}

\begin{document}

\title{Probabilistic Condition, Decision and Path Coverage of Circuit-based Quantum Programs}

\author{Daniel Fortunato}
\orcid{0000-0003-2596-6859}
\affiliation{%
  \institution{Faculty of Engineering of the University of Porto}
  \city{Porto}
  \country{Portugal}
}
\affiliation{%
  \institution{INESC-ID}
  \city{Lisboa}
  \country{Portugal}
}
\email{up202200549@edu.fe.up.pt}

\author{José Campos}
\orcid{0000-0001-7565-8382}
\affiliation{%
  \institution{Faculty of Engineering of the University of Porto}
  \city{Porto}
  \country{Portugal}
}
\affiliation{%
  \institution{LASIGE, Faculty of Science of the University of Lisbon}
  \city{Lisboa}
  \country{Portugal}
}
\email{jcmc@fe.up.pt}

\author{Rui Abreu}
\orcid{0000-0003-3734-3157}
\affiliation{%
  \institution{Faculty of Engineering of the University of Porto}
  \city{Porto}
  \country{Portugal}
}
\affiliation{%
  \institution{INESC-ID}
  \city{Lisboa}
  \country{Portugal}
}
\email{rui@computer.org}

\renewcommand{\shortauthors}{Fortunato et al.}

\begin{abstract}
Coverage criteria play a central role in assessing test adequacy in classical software, yet their effectiveness for quantum programs remains poorly understood and largely unexplored. In this paper, we propose six quantum-tailored criteria---condition, decision, and path coverage, and their probabilistic variants---adapted from their classical counterparts. We present \ourTool, a tool that computes these criteria for circuit-based quantum programs. We empirically evaluate these criteria on a large and diverse set of 540 circuits and analyze the coverage achieved. Our results show that while circuits frequently achieve high condition and decision coverage (97.56\% and 97.63\%, on average), path coverage remains limited (71.84\%), particularly in the presence of multi-controlled gates, which induce extreme path explosion and coverage imbalance. Moreover, to account for the probabilistic nature of quantum circuits, we introduce probabilistic coverage, which augments structural coverage with a confidence measure (88.87\%, 88.65\%, and 37.18\% for condition, decision, and path coverage, respectively, on average). Finally, through mutation testing, we find weak or no correlation between fault detection and structural coverage, consistent with observations in classical computing.
\end{abstract}
\maketitle


\section{Introduction}

Quantum computing leverages the principles of quantum mechanics and has emerged as a promising paradigm for performing computations beyond the reach of classical computers---for example, in chemistry~\cite{kais_introduction_2014} and materials science~\cite{yang_mixed-quantum-dot_2017}. By exploiting phenomena such as superposition and
entanglement~\cite{steane_quantum_1998, yanofsky_quantum_2008}, quantum algorithms can solve certain problems exponentially faster
than classical algorithms (e.g., Shor's algorithm~\cite{shor} for integer
factorization and Grover's algorithm~\cite{grover} for unstructured search). As quantum
hardware steadily advances~\cite{preskill_quantum_2018}, with increasing qubit counts and decreasing
error rates, developing correct and reliable quantum algorithms becomes increasingly
important. Ensuring correctness, however, is particularly challenging: quantum algorithms are fundamentally probabilistic,
and observing a quantum state (via measurement) collapses it, making intermediate states difficult to inspect without
disturbing execution. These unique aspects of quantum computation, combined with the relative immaturity of quantum programming toolchains, make \emph{testing and debugging} quantum algorithms significantly more difficult than in the classical domain.

In classical software development, \emph{testing} is an indispensable practice
for quality assurance~\cite{myers_art_2011}. Developers write and execute test
cases to assess whether the software's outputs or behaviors satisfy its specifications.
If they do not, a fault has been detected.
However, simply having test cases is not sufficient; developers must also assess how effective those tests are.
To that end, researchers have proposed \emph{testing criteria} (or \emph{adequacy criteria})~\cite{zhu_software_1997}
to evaluate and improve the thoroughness of test suites~\cite{shahid2011evaluation, fraser_software_2019}.

The most popular and widely used testing criteria in practice are \textbf{structural coverage criteria}~\cite{shahid_study_2011}
and \textbf{mutation testing}~\cite{jia_analysis_2011, petrovic_practical_2022}.
Structural coverage criteria (often called \emph{code coverage}) require tests to execute specific elements of a program's structure. For example,
\emph{line coverage} requires that each line of code be executed by at least one
test, whereas \emph{decision coverage} (commonly known as branch coverage)
requires that each possible branch (e.g., the \emph{then} and \emph{else} branches
of every \texttt{if} statement) be executed by at least one test.
Mutation testing takes a complementary approach. It intentionally injects
small faults (mutants) into the program and checks whether the tests fail on
the mutated versions. If the tests cannot detect a simple mutant, they may be
missing an important behavior.

To illustrate structural coverage, consider the following classical Python function that computes the maximum of two integers.
A thorough set of tests for this function should exercise both the case in which the first argument is larger and the case in which the second is larger.

\begin{wrapfigure}{l}{0.54\columnwidth}
\vspace{-1em}
\begin{tcolorbox}[enhanced, colback=white, boxrule=0pt, sharp corners, left=0pt, right=-10pt, top=-6pt, bottom=-10pt, frame hidden, interior hidden]
\begin{lstlisting}[language=Python, numbers=left]
def max(a: int, b: int) -> int:
    int: m;
    if a > b:
        m = a
    else:
        m = a # FAULT, it should be m = b
    return (m)
\end{lstlisting}
\end{tcolorbox}
\vspace{-1em}
\end{wrapfigure}

If we write a test that calls \texttt{max(5, 3)}, the function executes the
\texttt{then-branch} (line 3) and skips the \texttt{else}-branch (line 5); in doing so, it executes
lines 1, 2, 3, 4, and 7. Thus, this single test achieves line coverage of
$\frac{4}{6} = 67\%$. However, the decision coverage achieved by this test is
incomplete: the \texttt{else}-branch was never executed, meaning only one of the two
possible branches of the \texttt{if} statement was covered (i.e., $\frac{1}{2} = 50\%$ decision coverage).
When coverage falls below $100\%$, as in this example, developers are compelled to design additional tests to
explore alternative execution flows to increase coverage and confidence in the program under test.
For example, one could call \texttt{max(2, 4)} to trigger the \texttt{else}-branch (line 5)
and execute line 6. This additional test increases line and decision coverage to
$100\%$ and, most importantly, triggers the fault.
In classical software development, tools that report code coverage (e.g., JaCoCo~\cite{jacoco}
for Java and Coverage.py~\cite{coveragepy} for Python) are widely used to identify untested
parts of a program and to guide the creation of additional tests.
High coverage gives developers confidence that the tests have ``touched'' most of the
program's code, making it more likely to reveal faults in that code.
Besides completeness, coverage criteria have also been used in other testing
activities like regression testing~\cite{kazmi_effective_2018}, test
prioritization~\cite{srivastava2008test,Paterson2019}, test
generation~\cite{CAMPOS2018207,Rojas2015,CoverUp}, and test 
augmentation~\cite{santelices2008test,entbug,ctg}.

Quantum programs differ fundamentally from classical programs
because of their distinct mathematical foundations and probabilistic nature.
They are implemented as circuits that, when written in languages such as
OpenQASM~\cite{cross2017open} or Q\#~\cite{QDKDocs}
(or developed with frameworks such as Cirq~\cite{cirqDocument} and Qiskit~\cite{aleksandrowicz2019qiskit}),
consist of sequential operations (quantum gates and measurements) and lack the rich control-flow structures (loops, conditionals, and branches) found in classical programs. Thus, \textbf{classical coverage criteria are not directly applicable to quantum programs}.

For instance,
consider the \emph{Swap test}~\cite{barenco1997,buhrman2001}, a commonly used approach
in quantum software testing and debugging~\cite{miranskyy2025feasibilityquantumunittesting,9978402,9407042}. The Swap test estimates the squared
inner product between two quantum states, which indicates how similar they are~\cite{PMID:30992536}.
Values closer to $0$ indicate that the states are more orthogonal (and thus more different),
whereas values closer to $1$ indicate that they are more aligned (and thus more similar).

\begin{wrapfigure}{l}{0.39\columnwidth}
\vspace{-1em}
\begin{tcolorbox}[enhanced, colback=white, boxrule=0pt, sharp corners, left=0pt, right=-10pt, top=0pt, bottom=0pt, frame hidden, interior hidden]
\begin{lstlisting}[language=Python, numbers=left]
from qiskit import QuantumCircuit
qc = QuantumCircuit(3, 1)
qc.h([0])
qc.cswap([0], [1], [2])
qc.h([0])
qc.measure([0], [0])
\end{lstlisting}
\end{tcolorbox}
\vspace{-2em}
\caption{\emph{Swap test}~\cite{barenco1997,buhrman2001} written in Qiskit~\cite{aleksandrowicz2019qiskit}.}
\label{lst:swap-test-example}
\vspace{-1em}
\end{wrapfigure}

\Cref{lst:swap-test-example} provides an implementation of the Swap test written in Qiskit~\cite{aleksandrowicz2019qiskit}, the most popular framework for quantum computing~\cite{FERREIRA2025103217}.
First, it creates a circuit with three qubits and one classical bit to store the measurement result (line 2).
Then, it puts qubit \texttt{[0]} into superposition with a Hadamard gate (line 3) and
applies a controlled-swap gate (\texttt{cswap}, line 4) to qubits \texttt{[1]} and \texttt{[2]}, where
qubit \texttt{[0]} is the control. This operation swaps the states of \texttt{[1]} and
\texttt{[2]} if \texttt{[0]} is in the $\ket{1}$ state.
Finally, it applies a second Hadamard gate (line 5). At this point, if \texttt{[1]}
and \texttt{[2]} have equal states, then \texttt{[0]} will be in the $\ket{0}$ state.
However, if the states of \texttt{[1]} and \texttt{[2]} are orthogonal,
\texttt{[0]} will be in $\ket{0}$ or $\ket{1}$ with equal probability.
If we try to directly apply classical coverage metrics, such as line or
decision coverage, to this implementation of the Swap test, the outcome is
trivial. Running this (or any other quantum circuit) from start to finish executes
all lines exactly as written, because there are no alternate paths to take.
Therefore, every test 
inevitably achieves $100\%$ line coverage.
Likewise, decision coverage is meaningless in a straight-line circuit with no conditional branches,
i.e., there are no decision points where execution could diverge.
Even though quantum computation can explore multiple computational paths
simultaneously (thanks to superposition), those paths are not controlled by
branches that a test can trigger; they are inherent to the quantum
state of the circuit.

Thus, in this paper, we propose novel structural coverage criteria tailored to quantum circuits
that account for quantum controlled gates (e.g., \texttt{cx}, \texttt{cu}, and \texttt{cswap}), which determine whether some quantum computations are performed, similarly to \emph{if} statements in classical programs.
In a nutshell, the main contributions of this paper are as follows:
\begin{enumerate}[leftmargin=*]
  \item[\small{$\bigstar$}] Six \textbf{quantum-specific coverage criteria}---condition, decision, and path coverage, and their probabilistic variants---that capture quantum-circuit behavior beyond merely executing all instructions.

  \item[\small{$\bigstar$}] A \textbf{tool}, \ourTool, that automatically instruments quantum circuits written in OpenQASM, executes them, and computes these criteria.

  \item[\small{$\bigstar$}] An \textbf{empirical study} evaluating \ourTool's effectiveness and efficiency on 540 quantum circuits from MQT~Bench~\cite{quetschlich_mqt_2023}.
\end{enumerate}

\section{\ourTool}\label{sec:tool}

This section describes the novel coverage criteria and our tool, \ourToolFull\footnote{\ourTool homepage: \ourToolURL, accessed March 2026.}. \ourTool automatically instruments (\Cref{sec:tool:instr}), executes (\Cref{sec:tool:execution}), and computes coverage (\Cref{sec:tool:coverage,sec:tool:probabilisticcoverage}) for a given quantum circuit. It is a standalone Python tool that supports quantum circuits written in OpenQASM~\cite{cross2017open} (v2 and v3), and depends on Qiskit~\cite{aleksandrowicz2019qiskit} v2.1.1 and Qiskit Aer v0.17.1. 
Throughout this section, we use the Swap test circuit in \Cref{lst:swap-test-example} as a motivational example.

\subsection{Quantum Controlled Gates} \label{subsec:quantum_gates}

In quantum computing, many types of quantum gates perform distinct operations. In this work, we focus on quantum controlled gates~\cite{brylinski_universal_2002}. Controlled gates are multi-qubit gates that apply an operation (depending on the gate type) to a target qubit conditioned on the value of the control qubit(s). For instance, the Controlled-Not (\texttt{cx}) gate flips the target qubit if the control qubit is in the $\ket{1}$ state (similar to the classical XOR gate), and otherwise does nothing.
This means that a controlled gate can follow two different \emph{branches}: one where an operation is applied to the target qubit (the classical \emph{true} branch), and one where nothing happens (the \emph{false} branch).

To determine which \emph{branch} was taken, we could measure the control qubit(s) of each controlled gate immediately after execution. However, because the control qubit(s) might be entangled with other qubit(s), intermediate measurements would collapse the state of all entangled qubits and ruin the circuit's result. \Cref{sec:tool:instr} proposes a solution to bypass this constraint, at the cost of circuit efficiency (i.e., runtime).
At the time of writing, Qiskit v2.1.1 includes $22$ controlled gates: Single-Controlled Fixed (\texttt{cx}, \texttt{cy}, \texttt{cz}, \texttt{ch}, \texttt{csx}), Single-Controlled Parameterized (\texttt{crz}, \texttt{crx}, \texttt{cry}, \texttt{cu1}, \texttt{cu3}, \texttt{cp}), Single-Controlled Fixed Phase (\texttt{cs}, \texttt{csdg}), Multi-Controlled (\texttt{ccx}, \texttt{rccx}, \texttt{rcccx}, \texttt{c3sx}, \texttt{ccz}), General Controlled Unitary (\texttt{cu}), and Controlled Swap and Entangling (\texttt{cswap}, \texttt{dcx}, \texttt{ecr}).

Controlled gates such as \texttt{cx} have one control qubit, whereas others such as \texttt{ccx} have more than one (two in \texttt{ccx}'s case). Typically, the number of control qubits equals the number of ``c''s in the gate name. Except for \texttt{cx}, all controlled gates are internally decomposed (i.e., when transpiled) into one (or more) non-controlled unitary gates, zero (or more) non-controlled phase gates, and one (or more) \texttt{cx} gate(s). For example, the \texttt{cswap} gate in \Cref{lst:swap-test-example} is composed of five unitary/phase gates and seven \texttt{cx} gates.

\subsection{Instrumentation}\label{sec:tool:instr}

\begin{figure*}
  \centering
  \begin{subfigure}[t]{0.21\textwidth}
  \centering
\resizebox{\textwidth}{!}{
\Qcircuit @C=1.0em @R=0.2em @!R { \\
	 	\nghost{{q}_{0} :  } & \lstick{{q}_{0} :  } & \gate{\mathrm{H}} & \ctrl{1} & \gate{\mathrm{H}} & \meter & \qw & \qw\\
	 	\nghost{{q}_{1} :  } & \lstick{{q}_{1} :  } & \qw & \qswap & \qw & \qw & \qw & \qw\\
	 	\nghost{{q}_{2} :  } & \lstick{{q}_{2} :  } & \qw & \qswap \qwx[-1] & \qw & \qw & \qw & \qw\\
	 	\nghost{\mathrm{{c} :  }} & \lstick{\mathrm{{c} :  }} & \lstick{/_{_{1}}} \cw & \cw & \cw & \dstick{_{_{\hspace{0.0em}0}}} \cw \ar @{<=} [-3,0] & \cw & \cw\\
\\ }
}
  \vspace{-1em}\caption{Original circuit.  {\footnotesize \Cref{lst:swap-test-example} shows this circuit's Qiskit code.}}
  \label{fig:swap-test-example}
  \end{subfigure}
  \begin{subfigure}[t]{0.78\textwidth}
  \centering
\resizebox{\textwidth}{!}{
\Qcircuit @C=1.0em @R=0.2em @!R { \\
    \nghost{{q}_{0} :  } & \lstick{{q}_{0} :  } & \gate{\mathrm{H}} & \qw  & \qw & \qw & \ctrl{2}  & \qw & \qw & \qw  & \qw & \qw & \ctrl{2}  & \qw & \ctrl{1}  & \qw & \gate{\mathrm{P}\,(\mathrm{\frac{\pi}{4}})} & \ctrl{1}  & \qw & \qw & \qw  & \qw & \gate{\mathrm{H}} & \meter & \qw & \qw\\
    \nghost{{q}_{1} :  } & \lstick{{q}_{1} :  } & \gate{\mathrm{U}\,(\mathrm{\frac{\pi}{2},\frac{\pi}{2},\frac{-\pi}{2}})} & \ctrl{1} & \qw & \gate{\mathrm{U}\,(\mathrm{\frac{\pi}{2},\frac{-\pi}{2},\frac{\pi}{2}})} & \qw & \qw & \qw & \ctrl{1} & \qw & \gate{\mathrm{P}\,(\mathrm{\frac{\pi}{4}})} & \qw & \qw & \targ & \qw & \gate{\mathrm{P}\,(\mathrm{\frac{-\pi}{4}})} & \targ & \qw & \qw & \targ & \qw & \qw & \qw & \qw & \qw\\
    \nghost{{q}_{2} :  } & \lstick{{q}_{2} :  } & \gate{\mathrm{U}\,(\mathrm{\frac{\pi}{2},0,0.10})} & \targ & \qw & \gate{\mathrm{U}\,(\mathrm{1.47,\frac{-3\pi}{4},\frac{\pi}{2}})} & \targ & \qw & \gate{\mathrm{P}\,(\mathrm{\frac{\pi}{4}})} & \targ & \qw & \gate{\mathrm{P}\,(\mathrm{\frac{-\pi}{4}})} & \targ & \qw & \qw & \qw & \qw & \qw & \qw & \gate{\mathrm{U}\,(\mathrm{\frac{\pi}{2},0,\frac{-3\pi}{4}})} & \ctrl{-1} & \qw & \qw & \qw & \qw & \qw\\
    \nghost{\mathrm{{c} :  }} & \lstick{\mathrm{{c} :  }} & \lstick{/_{_{1}}} \cw & \cw & \cw & \cw & \cw & \cw & \cw & \cw & \cw & \cw & \cw & \cw & \cw & \cw & \cw & \cw & \cw & \cw & \cw & \cw & \cw & \dstick{_{_{\hspace{0.0em}0}}} \cw \ar @{<=} [-3,0] & \cw & \cw\\
\\ }
}
  \vspace{-1em}\caption{Transpiled circuit. {\footnotesize \Cref{lst:swap-test-example-transpiled} shows this circuit's Qiskit code.}}
  \label{fig:swap-test-example-transpiled}
  \vspace{1em}
  \end{subfigure}
  \begin{subfigure}[t]{\textwidth}
  \centering
\resizebox{\textwidth}{!}{
\Qcircuit @C=1.0em @R=0.2em @!R { \\
	 	\nghost{{q}_{0} :  } & \lstick{{q}_{0} :  } & \gate{\mathrm{H}} & \qw & \qw & \qw  & \qw & \qw & \ctrl{2} \qw & \gate{\includegraphics[height=1em]{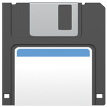}} & \qw  & \qw & \qw & \qw & \qw & \qw  & \qw & \qw & \ctrl{2} & \gate{\includegraphics[height=1em]{figures/floppy-crop.pdf}} & \qw  & \qw & \ctrl{1} & \gate{\includegraphics[height=1em]{figures/floppy-crop.pdf}} & \qw  & \qw & \gate{\mathrm{P}\,(\mathrm{\frac{\pi}{4}})} & \ctrl{1} & \gate{\includegraphics[height=1em]{figures/floppy-crop.pdf}} & \qw  & \qw  & \qw & \qw & \qw  & \gate{\includegraphics[height=1em]{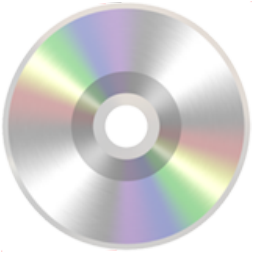}} & \gate{\mathrm{H}} & \meter & \qw & \qw\\
		\nghost{{q}_{1} :  } & \lstick{{q}_{1} :  } & \gate{\mathrm{U}\,(\mathrm{\frac{\pi}{2},\frac{\pi}{2},\frac{-\pi}{2}})} & \ctrl{1} & \gate{\includegraphics[height=1em]{figures/floppy-crop.pdf}} & \qw & \qw & \gate{\mathrm{U}\,(\mathrm{\frac{\pi}{2},\frac{-\pi}{2},\frac{\pi}{2}})} & \qw & \qw & \qw & \qw & \qw & \ctrl{1} & \gate{\includegraphics[height=1em]{figures/floppy-crop.pdf}} & \qw & \qw & \gate{\mathrm{P}\,(\mathrm{\frac{\pi}{4}})} & \qw & \qw & \qw & \qw & \targ & \qw & \qw & \qw & \gate{\mathrm{P}\,(\mathrm{\frac{-\pi}{4}})} & \targ & \qw & \qw & \qw & \qw & \targ & \qw & \qw & \qw & \qw & \qw & \qw\\
		\nghost{{q}_{2} :  } & \lstick{{q}_{2} :  } & \gate{\mathrm{U}\,(\mathrm{\frac{\pi}{2},0,0.10})} & \targ & \qw & \qw & \qw & \gate{\mathrm{U}\,(\mathrm{1.47,\frac{-3\pi}{4},\frac{\pi}{2}})} & \targ & \qw & \qw & \qw & \gate{\mathrm{P}\,(\mathrm{\frac{\pi}{4}})} & \targ & \qw & \qw & \qw & \gate{\mathrm{P}\,(\mathrm{\frac{-\pi}{4}})} & \targ & \qw & \qw & \qw & \qw & \qw & \qw & \qw & \qw & \qw & \qw & \qw & \qw & \gate{\mathrm{U}\,(\mathrm{\frac{\pi}{2},0,\frac{-3\pi}{4}})} & \ctrl{-1} & \gate{\includegraphics[height=1em]{figures/floppy-crop.pdf}} & \qw & \qw & \qw & \qw & \qw\\
		\nghost{\mathrm{{c} :  }} & \lstick{\mathrm{{c} :  }} & \lstick{/_{_{1}}} \cw & \cw & \cw & \cw & \cw & \cw & \cw & \cw & \cw & \cw & \cw & \cw & \cw & \cw & \cw & \cw & \cw & \cw & \cw & \cw & \cw & \cw & \cw & \cw & \cw & \cw & \cw & \cw & \cw & \cw & \cw & \cw & \cw & \cw & \dstick{_{_{\hspace{0.0em}0}}} \cw \ar @{<=} [-3,0] & \cw & \cw\\
\\ }
}
  \vspace{-1em}\caption{Instrumented circuit. {\footnotesize The \includegraphics[height=1em]{figures/cd-crop.pdf} and \includegraphics[height=1.1em]{figures/floppy-crop.pdf} boxes highlight the locations where the instrumentation injects additional instructions.  The former is used to compute decision coverage, and the latter is used to compute condition coverage of the original controlled gate (the \texttt{cswap} gate in this example).
  Each \includegraphics[height=1em]{figures/cd-crop.pdf} or \includegraphics[height=1em]{figures/floppy-crop.pdf} has a unique label and each is composed of two instructions available on Qiskit Aer's simulators: \texttt{save\_expectation\_value} which saves the expectation value of the corresponding qubit, and \texttt{save\_probabilities} which saves the measurement outcome probabilities vector of the corresponding qubit.  \Cref{lst:swap-test-example-instrumented} shows this circuit's Qiskit code.}}
  \label{fig:swap-test-example-instrumented}
  \end{subfigure}
  \vspace{-1em}
  \caption{Original, transpiled, and instrumented circuit of the Swap test written in Qiskit in \Cref{lst:swap-test-example}.}
  \label{fig:swap-test-example-circuits}
\end{figure*}

To be able to automatically compute the coverage of any controlled gate and therefore of any quantum circuit, \ourTool first performs two sequential transformations on the target quantum circuit: (1) transpilation of quantum controlled gates and (2) instrumentation of (1). Note that, typically, every quantum circuit has a ``control flow'', i.e., it contains at least one controlled gate; but in the off chance that it does not and the circuit is fully sequential, both these steps would be skipped, and, by definition, that circuit's coverage would be 100\%.

\subsubsection{Transpilation of quantum controlled gates}\label{sec:tool:instr:transpilation}
\ourTool transpiles all non-unitary controlled gates (e.g., \texttt{cswap}) to their
primitives~\cite{SetOfQuantumHardwarePrimitiveGatesInQiskit}, i.e., \texttt{u1}, \texttt{u2}, \texttt{u3}, \texttt{cx}, and \texttt{id}.
This is done to check the internal \texttt{cx} gates conditions of a given controlled gate to calculate our condition coverage criterion described in \Cref{sec:coverage:condition}.
This transpilation is performed by Qiskit Transpiler~\cite{QiskitTranspilerHomepage}.
The transpilation of \Cref{lst:swap-test-example} is shown in \Cref{lst:swap-test-example-transpiled}. We can see that the \texttt{cswap} gate (line 5) is removed and $17$ new gates are added (lines 6 to 22), of which seven are controlled gates \texttt{cx} (lines 8, 11, 13, 16, 17, 20, and 22). This step exposes all operations with a condition statement (i.e., \texttt{cx} gates), analogous to the classical \textit{if-else} statements. 

\begin{figure}
  \centering
  \begin{subfigure}[t]{0.39\columnwidth}
  \vspace{0pt}
\begin{tcolorbox}[enhanced, colback=white, boxrule=0pt, sharp corners, left=0pt, right=0pt, top=0pt, bottom=0pt, frame hidden, interior hidden]
\begin{lstlisting}[language=diff,numbers=left,basicstyle={\tiny\ttfamily}]
+ from math import pi
  from qiskit import QuantumCircuit
  qc=QuantumCircuit(3,3)
  qc.h([0])
- qc.cswap([0],[1],[2])
+ qc.u(pi/2,pi/2,-pi/2,[1])
+ qc.u(pi/2,0.0,0.10,[2])
+ qc.cx([1],[2])
+ qc.u(pi/2,-pi/2,pi/2,[1])
+ qc.u(1.47,-3*pi/4,pi/2,[2])
+ qc.cx([0],[2])
+ qc.p(pi/4,[2])
+ qc.cx([1],[2])
+ qc.p(pi/4,[1])
+ qc.p(-pi/4,[2])
+ qc.cx([0],[2])
+ qc.cx([0],[1])
+ qc.p(pi/4,[0])
+ qc.p(-pi/4,[1])
+ qc.cx([0],[1])
+ qc.u(pi/2,0.0,-3*pi/4,[2])
+ qc.cx([2],[1])
  qc.h([0])
  qc.measure([0],[0])
\end{lstlisting}
\end{tcolorbox}
\vspace{-1em}
\caption{Transpilation of the Qiskit code in \Cref{lst:swap-test-example}, i.e., decomposition of the \texttt{cswap} controlled gate. {\footnotesize \Cref{fig:swap-test-example-transpiled} shows the corresponding quantum circuit.  Lines of code added by the transpilation are \lightgreenhl{highlighted in green} and prefixed with \lightgreenhl{\texttt{+}}, and those removed are \lightredhl{highlighted in red} and prefixed with \lightredhl{\texttt{-}}.}\label{lst:swap-test-example-transpiled}}
\end{subfigure}
\hfill
  \begin{subfigure}[t]{0.58\columnwidth}
  \vspace{0pt}
\begin{tcolorbox}[enhanced, colback=white, boxrule=0pt, sharp corners, left=5pt, right=0pt, top=0pt, bottom=0pt, frame hidden, interior hidden]
\begin{lstlisting}[language=diff,numbers=left,basicstyle={\tiny\ttfamily}]
  from math import pi
  from qiskit import QuantumCircuit
+ from qiskit_aer import AerSimulator
+ from qiskit.quantum_info import Pauli
+ z=Pauli('Z')
  qc=QuantumCircuit(3,3)
  qc.h([0])
  qc.u(pi/2,pi/2,-pi/2,[1])
  qc.u(pi/2,0.0,0.10,[2])
  qc.cx([1],[2])
+ qc.save_exp_value(z,[1],'cswap_1_cx_1_value')
+ qc.save_prob([1],'cswap_1_cx_1_probability')
  qc.u(pi/2,-pi/2,pi/2,[1])
  qc.u(1.47,-3*pi/4,pi/2,[2])
  qc.cx([0],[2])
+ qc.save_exp_value(z,[0],'cswap_1_cx_2_value')
+ qc.save_prob([0],'cswap_1_cx_2_probability')
  qc.p(pi/4,[2])
  qc.cx([1],[2])
+ qc.save_exp_value(z,[1],'cswap_1_cx_3_value')
+ qc.save_prob([1],'cswap_1_cx_3_probability')
  qc.p(pi/4,[1])
  qc.p(-pi/4,[2])
  qc.cx([0],[2])
+ qc.save_exp_value(z,[0],'cswap_1_cx_4_value')
+ qc.save_prob([0],'cswap_1_cx_4_probability')
  qc.cx([0],[1])
+ qc.save_exp_value(z,[0],'cswap_1_cx_5_value')
+ qc.save_prob([0],'cswap_1_cx_5_probability')
  qc.p(pi/4,[0])
  qc.p(-pi/4,[1])
  qc.cx([0],[1])
+ qc.save_exp_value(z,[0],'cswap_1_cx_6_value')
+ qc.save_prob([0],'cswap_1_cx_6_probability')
  qc.u(pi/2,0.0,-3*pi/4,[2])
  qc.cx([2],[1])
+ qc.save_exp_value(z,[2],'cswap_1_cx_7_value')
+ qc.save_prob([2],'cswap_1_cx_7_probability')
+ qc.save_exp_value(z,[0],'cswap_1_value_1')
+ qc.save_prob([0],'cswap_1_probability_1')
  qc.h([0])
  qc.measure([0],[0])
\end{lstlisting}
\end{tcolorbox}
\vspace{-1em}
\caption{Instrumentation of the transpiled Qiskit code in \Cref{lst:swap-test-example-transpiled}. {\footnotesize \Cref{fig:swap-test-example-instrumented} shows the corresponding quantum circuit.  Lines of code added by the instrumentation are \lightgreenhl{highlighted in green} and prefixed with \lightgreenhl{\texttt{+}}.}
\label{lst:swap-test-example-instrumented}}
\end{subfigure}
\vspace{-1em}
\caption{Transpilation and instrumentation of the Qiskit code in \Cref{lst:swap-test-example}.\label{lst:swap-test-example-transpiled-and-swap-test-example-instrumented}}
\end{figure}

\subsubsection{Instrumentation of the transpiled quantum controlled gates}\label{sec:tool:instr:instrumentation}
As mentioned before, because intermediate measurements would disrupt the circuit's quantum state, we cannot measure the control qubit of every controlled gate mid-execution. To overcome this, for every \texttt{cx} gate in the transpiled circuit, \ourTool's instrumentation injects two instructions available on Qiskit Aer simulator:
\begin{enumerate}
  \item \texttt{save\_expectation\_value} which computes and stores the expectation value of one or more operators with respect to the quantum state during a simulation, i.e.,
  $
    \langle Z \rangle = \langle \psi \mid Z \otimes I \mid \psi \rangle
  $.
  It can be injected anywhere in the circuit; i.e., the simulator will record the expectation value at the injection point, and the circuit continues to run afterward.
  Recall that in real quantum hardware, if one measures a qubit in the Z-basis, the state collapses to $\ket{0}$ or $\ket{1}$.  In the Qiskit Aer simulator, \texttt{save\_expectation\_value} does not collapse the state.  Instead, it examines the full statevector at that point, applies the operator mathematically, and stores the resulting expectation value.
  Although this is a limitation of our approach, as the full statevector is only available in the simulation, we argue that our approach is the first to
  observe and reason about the control flow of quantum circuits. 
  \texttt{save\_expectation\_value} takes as input the operator (e.g., \texttt{Pauli('Z')}), the list of qubits the operator acts on, and a unique label name for later retrieving the stored result.
  $\langle Z \rangle$ ranges from $-1.0$ to $+1.0$ whereas:
  \begin{itemize}
    \item $-1.0$ means the control qubit of the controlled gate is certainly in the $\ket{1}$ state at that point in the circuit.
    \item $+1.0$ means the control qubit is certainly in the $\ket{0}$ state.
    \item $0.0$ means the control qubit is in a balanced superposition (or mixed state) with equal probability of being in the $\ket{0}$ and $\ket{1}$ state.
    \item Any other value (e.g., $-0.3$, $+0.7$) the qubit is in a state where measurement in the Z-basis is biased but not certain.
  \end{itemize}

  \item \texttt{save\_probabilities} 
  which computes and stores the measurement outcome probabilities vector of one or more operators with respect to the quantum state during a simulation.  
  It takes as input the list of qubits the operator acts on and a unique label name for retrieving the stored result.
\end{enumerate}
\Cref{lst:swap-test-example-instrumented} shows the instrumented circuit of the transpiled circuit in \Cref{lst:swap-test-example-transpiled}.

\subsection{Execution}\label{sec:tool:execution}

\begin{figure}
  \centering
  \begin{subfigure}[t]{0.78\columnwidth}
\begin{tcolorbox}[enhanced, colback=white, boxrule=0pt, sharp corners, left=0pt, right=0pt, top=0pt, bottom=0pt, frame hidden, interior hidden]
\begin{lstlisting}[language=Python]
from qiskit_aer import AerSimulator
simulator = AerSimulator(seed_simulator=0)
job = simulator.run(qc, shots=1024, seed_simulator=0)
result = job.result()

# Print the expected value of each cx's control qubit in cswap
print(result.data(0)['cswap_1_cx_1_value'])
print(result.data(0)['cswap_1_cx_2_value'])
print(result.data(0)['cswap_1_cx_3_value'])
print(result.data(0)['cswap_1_cx_4_value'])
print(result.data(0)['cswap_1_cx_5_value'])
print(result.data(0)['cswap_1_cx_6_value'])
print(result.data(0)['cswap_1_cx_7_value'])

# Print the expected probabilities of each cx's control qubit in cswap
print(result.data(0)['cswap_1_cx_1_probability'])
print(result.data(0)['cswap_1_cx_2_probability'])
print(result.data(0)['cswap_1_cx_3_probability'])
print(result.data(0)['cswap_1_cx_4_probability'])
print(result.data(0)['cswap_1_cx_5_probability'])
print(result.data(0)['cswap_1_cx_6_probability'])
print(result.data(0)['cswap_1_cx_7_probability'])

# Print the expected value of cswap's control qubit
print(result.data(0)['cswap_1_value_1'])
# Print the expected probabilities of cswap's control qubit
print(result.data(0)['cswap_1_probability_1'])
\end{lstlisting}
\end{tcolorbox}
\vspace{-1em}
\caption{Example of how one could execute the instrumented circuit in \Cref{lst:swap-test-example-instrumented} and get its intermediary values and their corresponding probabilities.\label{lst:swap-test-example-instrumented-execution}}
\end{subfigure}
\hfill
  \begin{subfigure}[t]{0.21\columnwidth}
\begin{tcolorbox}[enhanced, colback=white, boxrule=0pt, sharp corners, left=0pt, right=0pt, top=0pt, bottom=0pt, frame hidden, interior hidden]
\begin{lstlisting}[language=Python,numbers=none]







0.0
0.0
1.0
0.0
0.0
0.0
1.0



[0.5 0.5]
[0.5 0.5]
[1.0 0.0]
[0.5 0.5]
[0.5 0.5]
[0.5 0.5]
[1.0 0.0]


0.0


[0.5 0.5]
\end{lstlisting}
\end{tcolorbox}
\vspace{-1em}
\caption{Output of \Cref{lst:swap-test-example-instrumented-execution}.\label{lst:swap-test-example-instrumented-execution-output}}
\end{subfigure}
\vspace{-1em}
\caption{Execution of the instrumented circuit in \Cref{lst:swap-test-example-instrumented}.}
\label{lst:swap-test-example-instrumented-execution-and-swap-test-example-instrumented-execution-output}
\vspace{-10pt}
\end{figure}

To compute the values and corresponding probabilities of the control qubit of each controlled gate in the instrumented circuit, we execute the circuit using the Qiskit Aer simulator (lines 1--3 in \Cref{lst:swap-test-example-instrumented-execution}). Then, we retrieve the values and corresponding probabilities stored under each label---introduced by \texttt{save\_expectation\_value} and \texttt{save\_probabilities}, respectively---from the simulator's result (lines 4--22 in \Cref{lst:swap-test-example-instrumented-execution}). \Cref{lst:swap-test-example-instrumented-execution-output} reports the values computed by the simulator for the Swap test circuit. For example, lines 1 and 8 show that the control qubit of the first \texttt{cx} gate in the instrumented circuit was in a balanced superposition, with equal probability ($50\%$) of being in the $\ket{0}$ and $\ket{1}$ states. The same holds for the control qubit of the controlled gate \texttt{cswap} (lines 15 and 16). In contrast, lines 3 and 10 show that the control qubit of the third \texttt{cx} gate was in the $\ket{0}$ state with $100\%$ probability.

\subsection{Coverage}\label{sec:tool:coverage}

In this section, we formally define and describe the decision, condition, and path coverage criteria of quantum controlled gates. 

\smallskip\noindent\textbf{Definition 1:}
A quantum circuit $Q$ is composed of a collection of $M$ quantum controlled gates, $G(Q) = \{g_{1}, \dots, g_{i}, \dots, g_{M}\}$. In the Swap test circuit, $G(Q)$ only has one ($M$) controlled gate, i.e., $G(Q) = \{g_{1}\} = \{\text{\texttt{cswap}}\}$.

\smallskip\noindent\textbf{Definition 2:} 
A controlled gate $g \in G$ is a controlled unitary gate composed of a collection of $P$ qubits acting as controls, $L(g) = \{l_{1}, \dots, l_{k}, \dots, l_{P}\}$. In the Swap test circuit, the only controlled gate \texttt{cswap} ($g_{1}$) has one control qubit (\texttt{[0]}), i.e., $L(g_{1}) = \{l_{1}\} = [0]$.

\smallskip\noindent\textbf{Definition 3:}
Let $g^\text{true}$ and $g^\text{false}$ represent the two possible outcomes of a controlled gate $g \in G(Q)$.
$g^\text{true}$ and $g^\text{false}$ of a controlled gate $g \in G(Q)$ are:
\begin{itemize}
  \item $g^\text{true} = 1$ and $g^\text{false} = 0$, if and only if the expectation value of every control qubit $l \in L(g)$ is equal to $-1.0$.
  \item $g^\text{true} = 0$ and $g^\text{false} =1$, if and only if the expectation value of at least one control qubit $l \in L(g)$ is equal to $+1.0$.
  \item Both $g^\text{true}$ and $g^\text{false}$ equal to $1$, if and only if the expectation value of every control qubit $l \in L(g)$ is in $[-1.0, +1.0]$ and they do not satisfy one of the previous two scenarios.
\end{itemize}

\smallskip\noindent\textbf{Definition 4:}
The transpiled circuit of a controlled gate $g \in G(Q)$, $g'$, is composed of a collection of $N$ controlled-not gates \texttt{cx}, $C(g') = \{cx_{1}, \dots, cx_{j}, \dots, cx_{N}\}$. In the Swap test circuit example, $C(g'_{1})$ contains seven ($N$) controlled-not gates \texttt{cx}.

\smallskip\noindent\textbf{Definition 5:}
Let $cx^\text{true}$ and $cx^\text{false}$ represent the two possible outcomes of the control qubit in a controlled-not gate $cx \in C$.  $cx^\text{true}$ and $cx^\text{false}$ of a controlled-not gate $cx \in C$ are:
\begin{itemize}
  \item $cx^\text{true}$ = $1$ and $cx^\text{false}$ = $0$, if the expectation value of gate $cx$'s control qubit is $-1.0$.
  \item $cx^\text{true}$ = $0$ and $cx^\text{false}$ = $1$, if the expectation value of gate $cx$'s control qubit is $+1.0$.
  \item Both $1$ (true), if the expectation value of gate $cx$'s control qubit is in $]-1.0, +1.0[$.
\end{itemize}

\subsubsection{Decision Coverage}\label{sec:coverage:decision}
Also known as \emph{branch coverage}, ensures that every decision in a given program has been evaluated to both \texttt{true} and \texttt{false} at least once~\cite{fraser_software_2019}.  As already mentioned, controlled gates in quantum programs behave similarly to the classical conditional statement \texttt{if}.  Therefore, analyzing these conditional statements (i.e., controlled gates) reveals the program logic.  This can be relevant in potentially revealing faults and/or erroneous implementations.  We define the computation of the decision coverage of a given quantum circuit as:
\begin{equation}\label{eq:decision_coverage}\scalebox{0.95}{$\displaystyle
  \frac{\sum\limits_{g \in G(Q)}{\Big(g^\text{true} + g^\text{false}\Big)}}
  {2 \times |G(Q)|} \times 100\%$}
\end{equation}
As (a) there is only one controlled gate in the Swap test example (i.e., $G(Q) = \{g_{1}\} = \{\text{\texttt{cswap}}\}$ and $|G(Q)| = 1$), (b) \texttt{cswap} only has one control qubit ($L(g_{1}) = \{l_{1}\}$), and (c) the expected value of control qubit $l_{1}$ is $0.0$; its decision coverage is therefore:
\begin{equation*}\scalebox{0.95}{$\displaystyle
  \frac{\overbrace{1 + 1}^{g_{1}}}{2 \times 1} \times 100\% = 100\%$}
\end{equation*}
Recall Definition~3 for the computation of $g^{\text{true}}$ and $g^{\text{false}}$.

\subsubsection{Condition Coverage}\label{sec:coverage:condition}

Also known as \emph{predicate coverage}, ensures that each individual condition in a decision is evaluated to both \texttt{true} and \texttt{false} independently~\cite{fraser_software_2019}.  In this work, each controlled gate \texttt{cx} of a more complex controlled gate, e.g., \texttt{cswap}, is considered a condition.  We define the computation of the condition coverage of a given quantum circuit as:
\begin{equation}\label{eq:condition_coverage}\scalebox{0.95}{$\displaystyle
  \frac{\sum\limits_{g \in G(Q)}{~\sum\limits_{cx \in C(g')}{\Big(cx^\text{true} + cx^\text{false}\Big)}}}{2 \times \sum\limits_{g \in G(Q)}{|C(g')|}} \times 100\%$}
\end{equation}

As (a) there is only one controlled gate in the Swap test example (i.e., $G(Q) = \{g_{1}\} = \{\text{\texttt{cswap}}\}$, (b) \texttt{cswap} is transpiled to seven controlled-not gates $cx$ (i.e., $C(g'_{1}) = \{cx_{1}, cx_{2}, cx_{3}, cx_{4}, cx_{5}, cx_{6}, cx_{7}\}$ and $\sum_{g \in G(Q)}{|C(g'_{1})|} = 7$), (c) the expected value computed by the \texttt{save\_expectation\_value} instruction for $cx_{1}$, $cx_{2}$, $cx_{4}$, $cx_{5}$, and $cx_{6}$ is $0.0$ (see \Cref{lst:swap-test-example-instrumented-execution-output}), and (d) the expected value computed for $cx_{3}$ and $cx_{7}$ is $+1.0$; its condition coverage is therefore:
\begin{equation*}\scalebox{0.95}{$\displaystyle
  \frac{\overbrace{(1+1)}^{cx_{1}} + \overbrace{(1+1)}^{cx_{2}} + \dots + \overbrace{(1+1)}^{cx_{6}} + \overbrace{(1+0)}^{cx_{7}}}{2 \times 7} \times 100\% = 86\%$}
\end{equation*}
Recall Definition~5 for the computation of $cx^{\text{true}}$ and $cx^{\text{false}}$.

\subsubsection{Path Coverage}\label{sec:coverage:path}

This coverage criterion ensures that all different combinations of all conditions in a decision are exercised.  We define the computation of the path coverage of a given quantum circuit as: 

\begin{equation}\label{eq:path_coverage}\scalebox{0.95}{$\displaystyle
  \frac{\prod\limits_{g \in G(Q)} \prod\limits_{cx \in C(g')}\Big(cx^{\text{true}} + cx^{\text{false}}\Big)
  }{2^{\sum\limits_{g \in G(Q)}{|C(g')|}}} \times 100\%$}
\end{equation}

As (a) there is only one controlled gate in the Swap test example (i.e., $G(Q) = \{g_{1}\} = \{\text{\texttt{cswap}}\}$, (b) \texttt{cswap} is transpiled to seven controlled-not gates $cx$ (i.e., $C(g'_{1}) = \{cx_{1}, cx_{2}, cx_{3}, cx_{4}, cx_{5}, cx_{6}, cx_{7}\}$ and $\sum_{g \in G(Q)}{|C(g'_{1})|} = 7$), (c) the expected value computed by the \texttt{save\_expectation\_value} instruction for $cx_{1}$, $cx_{2}$, $cx_{4}$, $cx_{5}$, and $cx_{6}$ is $0.0$ (see \Cref{lst:swap-test-example-instrumented-execution-output}), and (d) the expected value computed for $cx_{3}$ and $cx_{7}$ is $+1.0$; its path coverage is therefore:
\begin{equation*}\scalebox{0.95}{$\displaystyle
  \frac{\overbrace{(1+1)}^{cx_{1}} \times \overbrace{(1+1)}^{cx_{2}} \times \dots \times \overbrace{(1+1)}^{cx_{6}} \times \overbrace{(1+0)}^{cx_{7}}}{2^7} \times 100\% = 25\%$}
\end{equation*}

\subsection{Probabilistic Coverage}\label{sec:tool:probabilisticcoverage}

Classic coverage reports, for example, $100\%$ decision coverage, assume both outcomes are executed at least once by the test cases.  Given the probabilistic nature of quantum circuits, however, there are different scenarios in which 100\% decision coverage is achieved.  For instance, for the Swap test, the decision coverage is $100\%$ if the control qubit of the controlled gate is in a balanced superposition, i.e., both decision outcomes are equally exercised. It is also $100\%$ if the control qubit is in $]-1.0, +1.0[~\setminus\{0\}$, i.e., one decision outcome is exercised more often than another.  To distinguish such cases, as the former represents an ideal scenario from a testing perspective~\cite{entbug} and the latter does not,
we combine coverage (as defined in \Cref{sec:tool:coverage}) with the Jain's fairness index~\cite{jain_quantitative_1984} of coverage's probabilities in a single value named \emph{probabilistic coverage}.

Jain's fairness index~\cite{jain_quantitative_1984} measures the fairness of resource allocation, ranging from $0$ to $1$.  It is widely used in computer science, particularly for analyzing network resource allocation~\cite{bin-obaid_fairness_2020, nowicki_more_2016}.  In this work, the index indicates perfect fairness when the index value is $1$ (i.e., all decision, condition, and path outcomes are equally exercised) and poor fairness when the value is closer to $0$. Moreover, it allows us to propose a \emph{probabilistic coverage} metric that represents the degree of confidence we have that the covered instructions (i.e., decisions, conditions, and paths) were, in fact, covered.

We now describe the computation of Jain's fairness index for each coverage criterion and the computation of the probabilistic coverage.

\smallskip\noindent\textbf{Definition 6:}
Let $g^\text{ptrue}$ and $g^\text{pfalse}$ represent the expected probability of the two possible outcomes of a controlled gate $g \in G(Q)$.
\begin{itemize}
  \item $g^\text{ptrue}$ is the sum of expected probabilities of all control qubits $l \in L(g)$ in which their expected value is in $]-1.0, +1.0]$.

  \item $g^\text{pfalse}$ is the sum of expected probabilities of all control qubits $l \in L(g)$ in which their expected value is in $[-1.0, +1.0[$.
\end{itemize}

\smallskip\noindent\textbf{Definition 7:}
Let $cx^\text{ptrue}$ and $cx^\text{pfalse}$ represent the probability of the possible outcomes of the control qubit in a controlled-not gate $cx \in C$.

\subsubsection{Probabilistic Decision Coverage}

The Jain's fairness index of the decision coverage is computed as:
\begin{equation}\label{eq:jain_fairness_index_for_decision_coverage}\scalebox{0.95}{$\displaystyle
  \frac{\bigg(\sum\limits_{g \in G(Q)}{g^{\text{ptrue}} + g^{\text{pfalse}}\bigg)^2}}{2 \times |G(Q)| \times \sum\limits_{g \in G(Q)}{\Big((g^{\text{ptrue}})^2 + (g^{\text{pfalse}})^2\Big)}} \times 100\%$}
\end{equation}
Given $g^{\text{ptrue}}$ and $g^{\text{pfalse}}$ of a controlled gate $g \in G(Q)$ always sum up to the number of control qubits of $G(Q)$, \Cref{eq:jain_fairness_index_for_decision_coverage} becomes:
\begin{equation}\label{eq:jain_fairness_index_for_decision_coverage_simplified}\scalebox{0.95}{$\displaystyle
  \frac{\bigg(\sum\limits_{g \in G(Q)}\Big|L(g)\Big|\bigg)^2}{2 \times |G(Q)| \times \sum\limits_{g \in G(Q)}{\Big((g^{\text{ptrue}})^2 + (g^{\text{pfalse}})^2\Big)}} \times 100\%$}
\end{equation}

As (a) there is only one controlled gate in the Swap test example (i.e., $G(Q) = \{g_{1}\} = \{\text{\texttt{cswap}}\}$), (b) $g_{1}$ only has one control qubit $L(g_{1}) = \{l_{1}\}$, and (c) the probability of the control qubit $l_{1}$ is $0.5$ for one outcome and $0.5$ for the other (according to the \texttt{save\_probabilities} instruction, see \Cref{lst:swap-test-example-instrumented-execution-output}).  Thus, the Jain's fairness index is:
\begin{equation*}\scalebox{0.95}{$\displaystyle
  \frac{1^2}{2 \times 1 \times \Big((0.5)^2 + (0.5)^2\Big)} \times 100\% = 100\%$}
\end{equation*}

Finally, the probabilistic decision coverage of a circuit $Q$ with $G$ controlled gates is:
\begin{equation}\label{eq:probabilistic_decision_coverage}
\begin{split}
  & \text{Decision Coverage of}~Q~\text{(\Cref{eq:decision_coverage})}~\times \\
  & \text{Jain's fairness index of the decision coverage of}~Q~\text{(\Cref{eq:jain_fairness_index_for_decision_coverage_simplified})}
\end{split}
\end{equation}
i.e., for the motivational example:$100\% \times 100\% = 100\%$.

\subsubsection{Probabilistic Condition Coverage}

The Jain's fairness index of the condition coverage is computed as:
\begin{equation}\label{eq:jain_fairness_index_for_condition_coverage}\scalebox{0.95}{$\displaystyle
   \frac{\bigg(\sum\limits_{g \in G(Q)}{\sum\limits_{cx \in C(g')}{\Big(cx^\text{ptrue} + cx^\text{pfalse}}\Big)\bigg)}^2}{2 \times \sum\limits_{g \in G(Q)}{|C(g')|} \times \sum\limits_{g \in G(Q)}{\sum\limits_{cx \in C(g')}{\Big((cx^\text{ptrue})^2 + (cx^\text{pfalse})^2\Big)}}} \times 100\%$}
\end{equation}
Given $cx^\text{ptrue}$ and $cx^\text{pfalse}$ of each gate $cx \in C$ always sum up to $1.0$, \Cref{eq:jain_fairness_index_for_condition_coverage} can be simplified to:
\begin{equation}\label{eq:jain_fairness_index_for_condition_coverage_simplified}\scalebox{0.95}{$\displaystyle
  \frac{\bigg(\sum\limits_{g \in G(Q)}{|C(g')|}\bigg)^2}{2 \times \sum\limits_{g \in G(Q)}{|C(g')|} \times \sum\limits_{g \in G(Q)}{\sum\limits_{cx \in C(g')}{\Big((cx^\text{ptrue})^2 + (cx^\text{pfalse})^2\Big)}}} \times 100\%$}
\end{equation}

In the motivation example, the controlled gate \texttt{cswap} transpiles to seven \texttt{cx} gates (i.e., $\sum_{g \in G(Q)}{|C(g')|} = 7$), and the values computed by the \texttt{save\_probabilities} instruction for the first, second, fourth, fifth, and sixth \texttt{cx} gate are $0.5$ ($cx^{ptrue}$) and $0.5$ ($cx^{pfalse}$), and $1.0$ ($cx^{ptrue}$) and $0.0$ ($cx^{pfalse}$) for the third and the seventh \texttt{cx} gate (see \Cref{lst:swap-test-example-instrumented-execution-output}).  Thus, the Jain's fairness index is:
\begin{equation*}\scalebox{0.95}{$\displaystyle
  \frac{7^2}{2 \times 7 \times \Big((0.5^2 + 0.5^2) + \dots + (1.0^2 + 0.0^2)\Big)} \times 100\% = 78\%$}
\end{equation*}

Finally, the probabilistic condition coverage of a circuit $Q$ with $G$ controlled gates is:
\begin{equation}\label{eq:probabilistic_condition_coverage}
\begin{split}
  & \text{Condition Coverage of}~Q~\text{(\Cref{eq:condition_coverage})}~\times \\
  & \text{Jain's fairness index of the condition coverage of}~Q~\text{(\Cref{eq:jain_fairness_index_for_condition_coverage_simplified})}
\end{split}
\end{equation}
i.e., for the motivational example: $86\% \times 78\% = 67\%$.

\subsubsection{Probabilistic Path Coverage}

The Jain's fairness index of the path coverage is computed as:

\begin{equation}\label{eq:jain_fairness_index_for_path_coverage}\scalebox{0.95}{$\displaystyle
   \frac{\bigg(\prod\limits_{g \in G(Q)} \prod\limits_{cx \in C(g')}\Big(cx^{\text{ptrue}} + cx^{\text{pfalse}}\Big)\bigg)^2}{2^{\sum\limits_{g \in G(Q)}{|C(g')|}} \times \prod\limits_{g \in G(Q)}{\prod\limits_{cx \in C(g')}{\Big((cx^\text{ptrue})^2 + (cx^\text{pfalse})^2\Big)}}} \times 100\%$}
\end{equation}
Given $cx^\text{ptrue}$ and $cx^\text{pfalse}$ of each gate $cx \in C$ always sum up to $1.0$, \Cref{eq:jain_fairness_index_for_path_coverage} can be simplified to:
\begin{equation}\label{eq:jain_fairness_index_for_path_coverage_simplified}\scalebox{0.95}{$\displaystyle
  \frac{1}{2^{\sum\limits_{g \in G(Q)}{|C(g')|}} \times \prod\limits_{g \in G(Q)}{\prod\limits_{cx \in C(g')}{\Big((cx^\text{ptrue})^2 + (cx^\text{pfalse})^2\Big)}}} \times 100\%$}
\end{equation}

In the motivation example, the controlled gate \texttt{cswap} transpiles to seven \texttt{cx} gates (i.e., $\sum_{g \in G(Q)}{|C(g')|} = 7$), and the values computed by the \texttt{save\_probabilities} instruction for the first, second, fourth, fifth, and sixth \texttt{cx} gate are $0.5$ ($cx^{ptrue}$) and $0.5$ ($cx^{pfalse}$), and $1.0$ ($cx^{ptrue}$) and $0.0$ ($cx^{pfalse}$) for the third and the seventh \texttt{cx} gate (see \Cref{lst:swap-test-example-instrumented-execution-output}).  Thus, the Jain's fairness index is:
\begin{equation*}\scalebox{0.95}{$\displaystyle
  \frac{1}{2^{7} \times \Big((0.5^2 + 0.5^2) \times \dots \times (1.0^2 + 0.0^2)\Big)} \times 100\% = 25\%$}
\end{equation*}

Finally, the probabilistic path coverage of a program $Q$ with $G$ controlled gates is:
\begin{equation}\label{eq:probabilistic_path_coverage}
\begin{split}
  & \text{Path Coverage of}~Q~\text{(\Cref{eq:path_coverage})}~\times \\
  & \text{Jain's fairness index of the path coverage of}~Q~\text{(\Cref{eq:jain_fairness_index_for_path_coverage_simplified})}
\end{split}
\end{equation}
i.e., for the motivational example: $25\% \times 25\% = 6.25\%$.

\section{Empirical Study}\label{sec:study}

In this section, we investigate the efficiency of \ourTool and the effectiveness of default inputs in exercising quantum programs' structure. In a nutshell, we address the following research questions:
\begin{enumerate}[label=\textbf{RQ\arabic*:}]
  \item How long does \ourTool take to instrument a circuit?
  \item What is the overhead of \ourTool's instrumentation? 
  \item How effective are $\ket{0}$ inputs at exercising controlled gates?
  \item How does structural coverage correlate with fault detection?
\end{enumerate}

\subsection{Experimental Subjects}

To the best of our knowledge, there are three sets of subjects, i.e., quantum circuits, suitable for our empirical study:
VeriQBench~\cite{chen_veriqbench_2022}, QASMBench~\cite{li_qasmbench_2023}, and MQT~Bench~\cite{quetschlich_mqt_2023}. They contain 22, 43, and 29 quantum algorithms, respectively, and each provides multiple variants per algorithm with different numbers of qubits. In total, VeriQBench contains 966 OpenQASM circuits, QASMBench 132, and MQT~Bench 1,943.

However, only MQT~Bench simultaneously provides a diverse set of circuits that satisfy \ourTool's requirements (i.e., circuits that can be loaded with Qiskit and contain at least one quantum controlled gate) and has been used in similar studies~\cite{hopf_improving_2025,apak_ketgpt_2024,quetschlich_compiler_2023,UsandizagaMutation2025}.
First, 123 circuits in VeriQBench and nine circuits in QASMBench cannot be loaded with Qiskit~\cite{VeriQIssue5,VeriQIssue6,QASMBenchIssue9}.
All circuits in MQT~Bench can be loaded with Qiskit.
Second, 714 out of 966 circuits in VeriQBench contain at least one quantum controlled gate (implementing 19 out of 22 algorithms, 86.37\%), and 117 out of 132 in QASMBench (36 out of 43 algorithms, 83.72\%).
By contrast, 1,912 out of 1,943 circuits in MQT~Bench contain at least one quantum controlled gate (implementing 27 out of 29 algorithms, 93.10\%). Quantum algorithms \texttt{grover-\{noancilla$|$v-chain\}} with two qubits (two circuits) and all variants of \texttt{portfolioqaoa} and \texttt{qaoa} (i.e., $14 + 15 = 29$ circuits) lack controlled gates.
In summary, MQT~Bench provides by far the largest number of usable circuits for our study (1,912 vs.\ 714 in VeriQBench and 117 in QASMBench) and high algorithm-level coverage among circuits with controlled-gate behavior (27/29 algorithms vs.\ 19/22 in VeriQBench and 36/43 in QASMBench).

\subsubsection{MQT~Bench's circuits analysis}

Overall, the most frequently used quantum controlled gate in MQT~Bench is \texttt{cx} (49,062 occurrences; average of 2829.8 and median of 73.0), followed by \texttt{cu1} (48,966 occurrences; average of 152.5 and median of 0.0) and \texttt{cp} (25,575 occurrences; average of 992.8 and median of 0.0). Although controlled gates such as \texttt{cu} and \texttt{c3sx} are available in Qiskit and supported by \ourTool, they are not used in any circuit.
The \texttt{cx} gate is also the most prevalent across circuits: 793 circuits use only \texttt{cx}, and 1,375 out of 1,943 circuits use it in combination with other controlled gates.

\subsection{Experimental Setup}\label{sec:exp_setup}

All experiments were conducted on a machine running Ubuntu 22.04.5 LTS with Linux kernel version 5.15.0-119-generic. The running environment consisted of two Intel® Xeon® E5-2640 v2 CPUs and 16GB of DDR3 RAM clocked at 1600 MHz.  The total computational cost for all experiments was $\approx145$~CPU-years.

\subsection{Experimental Methodology}\label{sec:exp_procedure}
To answer our RQs, our empirical study followed the procedure described below.

\subsubsection*{RQ1}
We first transpiled all quantum controlled gates in each circuit to their hardware primitives (see \Cref{sec:tool:instr:transpilation}) using optimization level 0 and a fixed random seed (i.e., 0), ensuring a consistent and reproducible process across all circuits. Next, we instrumented each transpiled controlled gate as described in \Cref{sec:tool:instr:instrumentation}. We measured, in seconds, the time required to transpile and instrument each circuit. Additionally, we computed Spearman's rank correlation coefficient between runtime and circuit metadata (i.e., number of qubits, number of controlled gates, depth, and size).

In RQ1, we used all quantum circuits with at least one quantum controlled gate (i.e., 1,912 circuits out of 1,943) to assess \ourTool's instrumentation runtime.

\subsubsection*{RQ2}
We executed each original quantum circuit and its instrumented version 10 times on the Qiskit Aer simulator using the same backend configuration: a single shot (because our goal is to measure execution-time overhead rather than outcome distributions) and a fixed random seed (0). For each execution, we measured the execution time (in seconds) of both circuits (original and instrumented) and discarded 20\% of the repetitions with the highest and lowest time values to reduce hardware bias~\cite{PEREIRA2021102609} (e.g., CPU warm-up effects). For both circuits, we also measured the number of controlled gates, depth (i.e., number of instructions), and size (in kilobytes).
We used differences in execution time, number of controlled gates, depth, and size between the original and instrumented circuits to assess the overhead introduced by \ourTool. Furthermore, we used the Vargha-Delaney $\hat{A}_{12}$ effect size to assess whether overhead exists, and the Wilcoxon-Mann-Whitney U-test (with a 95\% confidence level) to evaluate whether any observed overhead is statistically significant.

In RQ2, we used all quantum circuits with at least one quantum controlled gate (i.e., 1,912 circuits out of 1,943) to assess \ourTool's instrumentation overhead regarding depth and size of the instrumented circuits (both static metrics that do not require circuit execution), and 540 circuits to assess \ourTool's runtime overhead when executing instrumented circuits. Given our hardware constraints (i.e., 16GB of RAM), we cannot execute circuits with more than 32 qubits in the quantum simulator.

\subsubsection*{RQ3}
We executed each instrumented circuit with its default inputs (i.e., all qubits initialized with $\ket{0}$) on the Qiskit Aer simulator using a single shot and a fixed random seed (0) (see \Cref{sec:tool:execution}). A single shot is sufficient because our computation relies on \texttt{save\_expectation\_value}, which is derived from the simulated quantum state for a given input rather than estimated from repeated measurement sampling; therefore, additional shots would only increase runtime without changing these expectation values. Second, we computed the structural coverage criteria defined in \Cref{sec:tool:coverage,sec:tool:probabilisticcoverage} for each instrumented circuit.
In RQ3, we used the instrumented versions of the 540 quantum circuits that we can execute in the simulator.

\subsubsection*{RQ4}

We first computed the statevector, i.e., a mathematical representation of the complete quantum state system, of each original quantum circuit.

Second, to simulate faults that developers might introduce, we applied the three commonly used quantum mutation operators~\cite{li2026methodologicalanalysisempiricalstudies}, i.e., QGR, QGD, and QGI, one at a time to the original quantum circuit, as proposed by \citet{fortunato_mutation_2022} and \citet{mendiluze_muskit_2021}.
Interested readers can find definitions and examples of each mutation operator in~\cite{fortunato_mutation_2022}.

Third, we computed the statevector of each mutated circuit and compared it with the statevector of the original circuit. Although others have used different procedures to assess the fate of each mutant (e.g., comparing outputs of original and mutated circuits, or comparing their probabilities with statistical tests~\cite{ali_assessing_2021,mendiluze_muskit_2021,wang_quito_2021,UsandizagaMutation2025}), \citet{miranskyy2025feasibilityquantumunittesting} empirically demonstrated that statevector comparison is more effective.

Fourth, we classified each mutated circuit as follows: \emph{killed}, if the statevector of the original circuit was not equivalent\footnote{We used the \texttt{equiv} method available in Qiskit, \url{https://quantum.cloud.ibm.com/docs/en/api/qiskit/qiskit.quantum\_info.Statevector\#equiv} accessed March 2026, with a tolerance value of $1 \times 10^{-8}$ for comparing any two statevectors.} to the statevector of the mutated circuit; \emph{survived}, if the statevector of the original circuit was equivalent to the statevector of the mutated circuit; and \emph{timeout}, if the execution time of the mutated circuit was 10\% higher than the execution time of its original circuit. We then calculated \emph{mutation score} as the ratio of killed mutants to the total number of generated mutants (including timeouts) for each quantum circuit.

Finally, we computed Spearman's rank correlation coefficient to measure the strength and direction of the relationship between mutation score and the structural coverage criteria defined in \Cref{sec:tool:coverage,sec:tool:probabilisticcoverage}.\footnote{Because our data is not normally distributed, we used Spearman's rank correlation coefficient instead of, for example, the well-known Pearson correlation coefficient.}

Moreover, because long circuits with many qubits and gates would (a) produce millions of mutated circuits and (b) require substantial time to compute each mutant statevector (recall that the time complexity of statevector computation is $O(|G| \times 2^n)$, where $|G|$ is the number of gates and $n$ is the number of qubits), we set a hard timeout of 24 hours for the RQ4 procedure. Thus, in RQ4, we used 248 quantum circuits (with qubit counts in $[2, 20]$), i.e., those for which it was possible to complete the steps described above within 24 hours.

\subsection{Threats to Validity}\label{sec:evaluation:threats}

Based on the guidelines reported by \citet{yin2009case} and \citet{wohlin2012experimentation}, we have taken all reasonable steps to mitigate the effect of potential
threats, which are described in detail in this subsection.

\subsubsection{External Validity}

First, our subjects (all from MQT~Bench and all quantum algorithms) may not represent the full diversity of real-world quantum software (e.g., proprietary or domain-specific systems), which limits generalization. Given that quantum software engineering is still in its infancy, to the best of our knowledge no other dataset offers a similarly large and diverse set of circuits. 

Second, due to hardware limitations, we selected a subset of the 1,943 circuits in MQT~Bench. Even so, we included at least one circuit for each of its 29 quantum algorithms.

Third, the coverage reported in RQ3 and the mutation score in RQ4 are based on default inputs (i.e., all qubits in the $\ket{0}$ state). Other input states might yield different coverage values. As future work, we plan to evaluate additional input states, including those 
generated by approaches such as QuraTest~\cite{ye_quratest_2023} and NovaQ~\cite{NovaQ}.

Fourth, the fault types used in RQ4 might be unrepresentative of real faults in quantum software. This threat is hard to mitigate because large datasets of real quantum faults are scarce: QBugs~\cite{QBugs} is not available, and Bugs4Q~\cite{ZHAO2023111805} contains only 20 source-code \emph{bugs} (out of 42), most of which could be mimicked by the mutation operators we used. 

\subsubsection{Internal Validity}

We carried out our experiments using the Qiskit transpiler and Qiskit Aer simulator. Although our instrumentation and execution process was consistent, randomness and limitations in handling mid-circuit operations might introduce subtle biases. Our process was automated, but defects or edge cases are still possible. To mitigate these risks, we manually validated outputs for selected examples and used unit tests to verify that the pipeline was correctly implemented.

\section{Results}\label{sec:results}

In this section, we present the results from the experimental procedure and the answers to the proposed RQs. 

\subsection{Results and Answer to RQ1}

On average, \ourTool took 28.56 seconds to instrument a quantum circuit (median of 8.77), and 1,298 circuits (67.89\%) were instrumented in less than 28.56 seconds.  The longest instrumentation performed by \ourTool took 590.23 seconds ($\approx10$ minutes) for the \texttt{qwalk-noancilla} (with 13 qubits) circuit, which has a particularly high circuit depth (244,927) and size (65,307.697 KB), and it is the circuit with the highest number of controlled gates (147,030).

Moreover, we also computed the correlation of \ourTool's instrumentation runtime vs.\ number of qubits, number of controlled gates, depth, and size of a circuit. According to Spearman correlation coefficients, there is a statistically significant strong and positive correlation between \ourTool's instrumentation runtime and number of qubits ($\rho=0.61$); and a statistically significant very strong and positive correlation between \ourTool's instrumentation runtime and number of controlled gates, depth, and size ($\rho=0.98$, $\rho=0.82$, and $\rho=0.91$, respectively).
Overall, these results are to be expected.  Although we could have a circuit with a high depth or size that contains no controlled gates, this is not the case in our set of subjects. Note that if we tried to instrument a circuit with no controlled gates, this would be instantaneous, as there would be no change to the circuit.  Deeper and larger circuits contain many controlled gates, which result in longer instrumentation times due to the overhead of inserting multiple \texttt{save\_expectation\_value} and \texttt{save\_probabilities} instructions per controlled gate.

\smallskip\noindent\textbf{RQ1 (answer):} \ourTool takes, on average, 28.56 seconds to instrument a circuit, and it is capable of instrumenting 983 circuits (51.41\%) in less than 10 seconds and 1,609 circuits (84.15\%) within one minute.

\subsection{Results and Answer to RQ2}

Overall, there is no increase in the number of qubits, as \ourTool's instrumentation does not add any, and there is no increase in the depth of the circuit, as the instrumented code is not part of the circuit itself. There is, however, a statistically significant increase in the size of the circuit ($\hat{A}_{12}=0.32$ and $p$-value $< 0.0001$) and runtime ($\hat{A}_{12}=0.55$ and $p$-value $0.0049$). On average, the instrumented circuits are +4.66 times larger than the original circuits, 3,164.78 (median of 916.98 KB) vs.\ 678.77 KB (median of 185.82 KB); and take 11.85 times more time to run, 2002.95 seconds (but only a median of 0.90) vs.\ 169.00 seconds (median of just 0.92). Worth pointing out that 429 instrumented circuits (79.44\% out of 540) run in less than one minute.
Nevertheless, this means there is a small number of extremely time-consuming circuits that pull the average far above the common experience.  For instance, the top-3 time-consuming circuits are \texttt{qwalk-v-chain} (with 31 qubits) 79,086.56 seconds, \texttt{random} (with 32 qubits) 74,359.36 seconds, and \texttt{qnn} (with 32 qubits) 67,457.87 seconds (21.97, 20.66, and 18.74 hours, respectively).  These account for 21.36\% of the total runtime of the 540 circuits, while representing only 0.56\% of the circuits. These are, first, among the circuits with the most qubits and secondly, with the most controlled gate occurrences.

\smallskip\noindent\textbf{RQ2 (answer):} Quantum circuits instrumented by \ourTool are +4.66 times larger than the original circuits and take 11.85 times more time to run.  Nevertheless, 429 instrumented circuits (79.44\% out of 540) run in less than one minute.
Thus, the overhead is negligible for most circuits, but becomes more significant for circuits with a high number of qubits, 
gates, and depth.

\subsection{Results and Answer to RQ3}

\Cref{tab:coverage-default-singlecol} shows the obtained coverage values in three different groups: 
(i) ``All'' circuits, i.e., 540,
(ii) ``Non-complex'' circuits, i.e., circuits with only single-controlled gates (252),
and (iii) ``$\geq$1 complex'' circuits, i.e., circuits with at least one multi-controlled gate (288).  

\begin{table}
\centering
\small
\caption{Coverage values grouped by control-flow complexity.\label{tab:coverage-default-singlecol}}
\setlength{\tabcolsep}{3pt}
\vspace{-1em}\resizebox{\columnwidth}{!}{\begin{tabular}{ll
                rrrr
                rrrr
                rrrr}
\toprule
 &  &
 \multicolumn{4}{c}{\textbf{Coverage (\%)}} &
 \multicolumn{4}{c}{\textbf{Jain (\%)}} &
 \multicolumn{4}{c}{\textbf{Probabilistic (\%)}} \\
\cmidrule(lr){3-6}
\cmidrule(lr){7-10}
\cmidrule(lr){11-14}
\textbf{Metric} & \textbf{Group}
 & \textbf{Min} & \textbf{Max} & \textbf{Median} & \textbf{Avg}
 & \textbf{Min} & \textbf{Max} & \textbf{Median} & \textbf{Avg}
 & \textbf{Min} & \textbf{Max} & \textbf{Median} & \textbf{Avg} \\
\midrule
\multirow{3}{*}{Condition}
 & All
 & 50.00 & 100.00 & 100.00 & 97.56
 & 50.00 & 100.00 & 93.32 & 90.72
 & 25.00 & 100.00 & 92.27 & 88.87 \\
 & Non-complex
 & 75.00 & 100.00 & 100.00 & 99.29
 & 51.67 & 100.00 & 92.98 & 90.92
 & 50.00 & 100.00 & 92.98 & 90.40 \\
 & $\geq$1 complex
 & 50.00 & 100.00 & 99.62 & 95.58
 & 50.00 & 100.00 & 93.36 & 90.48
 & 25.00 & 100.00 & 90.57 & 87.12 \\
\midrule
\multirow{3}{*}{Decision}
 & All
 & 50.00 & 100.00 & 100.00 & 97.63
 & 50.00 & 100.00 & 92.61 & 90.45
 & 25.00 & 100.00 & 91.33 & 88.65 \\
 & Non-complex
 & 75.00 & 100.00 & 100.00 & 99.29
 & 51.67 & 100.00 & 92.98 & 90.92
 & 50.00 & 100.00 & 92.98 & 90.40 \\
 & $\geq$1 complex
 & 50.00 & 100.00 & 100.00 & 95.74
 & 50.00 & 100.00 & 92.15 & 89.90
 & 25.00 & 100.00 & 88.93 & 86.65 \\
\midrule
\multirow{3}{*}{Path}
 & All
 & 0.00 & 100.00 & 100.00 & 71.84
 & 0.00 & 100.00 & 0.10 & 37.55
 & 0.00 & 100.00 & 0.02 & 37.18 \\
 & Non-complex
 & 1.56 & 100.00 & 100.00 & 91.11
 & 0.00 & 100.00 & 0.61 & 37.35
 & 0.00 & 100.00 & 0.23 & 36.99 \\
 & $\geq$1 complex
 & 0.00 & 100.00 & 25.00 & 49.82
 & 0.00 & 100.00 & 0.00 & 37.78
 & 0.00 & 100.00 & 0.00 & 37.40 \\
\bottomrule
\end{tabular}}
\end{table}

\subsubsection{Coverage}
On average, 97.56\% ($\sigma=5.45$), 97.63\% ($\sigma=5.40$), and 71.84\% ($\sigma=42.73$) of all conditions, decisions, and paths, respectively, of all controlled gates are covered.  Standard deviation values suggest that condition and decision coverage are concentrated around the average, whereas path coverage shows greater variability.  On one side, two circuits achieved the lowest condition and decision coverage (i.e., 50\%) and eight circuits the lowest path coverage (i.e., a value close to 0).  On the other side, 368 circuits (out of 540, 68.15\%) achieved the highest condition, decision, and path coverage. 
\emph{How could one single input achieve 100\% on all criteria?}

The \texttt{ae} circuit (with five qubits) achieved full coverage on all criteria.  This circuit starts by applying rotation gates to all the circuit's qubits, i.e., putting qubits in a superposition state. 
This ensures all control flow paths are executed, thereby achieving maximum coverage.
We observed the opposite in \texttt{qpeinexact} (with two qubits), only 50\% condition and decision coverage.  This circuit does not start by applying any rotation gates to its qubits, so they remain in their initial state of 0 or 1 (i.e., not in a superposition state).

At the group level, and as expected, the condition and decision coverage is equal in group (ii) as each decision is composed of only one condition.  In classical computing, this would translate to a single \texttt{if} statement (i.e., the decision) with one single condition to satisfy.
In group (iii), the average decision and condition coverage is slightly different, 95.74 vs.\ 95.58\%.  Path coverage is lower in group (iii), less than 50\%, than in (ii), 91.11\%. 
There are fewer paths to execute in group (ii) than in (iii), $3.22 \times 10^{9}$ vs.\ $2.46 \times 10^{114}$.

\subsubsection{Probabilistic Coverage}

Given the probabilistic nature of quantum computing, the structural coverage of a quantum circuit is associated with a probability (as opposed to classical computing), i.e., the degree of confidence that executed decisions, conditions, and paths were, in fact, executed.  In the quantum realm, we might have 100\% decision, or condition, or path coverage with a low or high probability associated.  To distinguish such cases, since a coverage value associated with a high probability value represents the ideal testing scenario, and the opposite does not, we proposed \emph{probabilistic coverage} criteria (see \Cref{sec:tool:probabilisticcoverage}) that combine coverage values with Jain's fairness index.

On average, the Jain's fairness index of decision, condition, and path coverage is similar in all three groups.  As for condition and decision coverage, the standard deviation of Jain's fairness index values in group (i) ($\sigma=10.36$ and $\sigma=10.24$) suggests that they are concentrated around the average, while path coverage ($\sigma=47.47$) exhibits high variability.  On the one hand, two circuits achieved a Jain's fairness index of 50\% for condition and decision coverage, and 17\% for path coverage.  On the other hand, 190 circuits (out of 540, 35.19\%) achieved a Jain's fairness index of 100\% for condition, decision, and path coverage.
\emph{In which scenarios does a circuit achieve a Jain's fairness index of 100\%?}

Similarly to coverage, starting by putting qubits into superposition (e.g., via Hadamard gates) could lead to a high Jain index for coverage criteria.  For example, the \texttt{ae} (with five qubits) circuit, which achieves 100\% decision, condition, and path coverage, also achieves a Jain's index of 100\% for each coverage criterion.  Thus, one could say that the probabilistic decision, condition, and path coverage of this example would be 100\%.  There are 168 circuits (out of 540, 31.11\%) that achieved 100\% condition, decision, and path coverage but did not have a Jain's index of 100\%.
For instance, the circuit \texttt{routing} (with two qubits) achieved 100\% decision, condition, and path coverage, but achieved a Jain's index and probabilistic condition and decision coverage of 51.67\%, and a Jain's index and probabilistic path coverage of 13.80\%.
Note: The lowest coverage values occur almost exclusively in circuits with multi-controlled gates, whereas the highest coverage values are predominantly observed in circuits with only non-complex controlled gates.  These results highlight both the strengths and the limitations of default inputs when applied to quantum circuits with varying control complexity.

\smallskip\noindent\textbf{RQ3 (answer):} 
Regarding coverage criteria, default inputs cover, on average, 97.63\% of all decisions, 97.56\% of all conditions, and 71.84\% of all paths.  Regarding probabilistic coverage, default inputs achieved 88.65\%, 88.87\%, and 37.18\%.

\subsection{Results and Answer to RQ4}

To assess whether coverage is correlated with fault detection, we conducted a mutation analysis experiment by applying three mutation operators
(see \Cref{sec:exp_procedure}) to 248 circuits. 
Then, we compared the correlation between the mutation score and the coverage criteria. 

\subsubsection{Mutation analysis}
.
Overall, we generated 287,307 mutants of which 271,845 (94.62\%) were detected (i.e., \emph{killed}).  14,604 mutants survived (5.08\%) and 858 timeout (0.30\%).

The high number of kills can be explained by the use of the circuit's statevector to assess whether a mutated circuit is equivalent to its original circuit.  In previous works (e.g., ~\cite{mendiluze_muskit_2021,fortunato_mutation_2022,wang_mutation-based_2022}), others have compared the original circuit's outputs with the mutant circuit's outputs by comparing output measurements and their distributions~\cite{ali_assessing_2021}.  Quantum mutations are sometimes negligible, to the point that they might neither affect outputs nor distributions and thus go undetected during testing.  However, this is not the case when assessing whether circuit A's statevector is equivalent to circuit B's statevector, as any change to the circuit will affect it. 

Regarding survived mutants, and given statevector's comparison effectiveness~\cite{miranskyy2025feasibilityquantumunittesting}, it is likely that the surviving mutants are equivalent, i.e., mutated circuits that behave as the original circuits.  We leave this analysis for future work.  All generated mutants and mutants' data are available in the accompanying supplementary material and could be explored and analyzed by others.

Regarding timeouts, we observed that the number of timeouts is higher in QGI (0.28\%; 376 out of 134,036) and QGR operators (0.34\%; 456 out of 134,036) than in QGD (0.14\%; 26 out of 19,235). The QGI operator increases the circuit size by adding a single quantum gate. The QGR operator, which replaces a quantum gate with a syntactically equivalent one, might replace a shorter-executing gate with a longer-executing gate. We suggest, for future work, investigating the execution time of each quantum gate and integrating that information into the QGR operator or a novel one. Finally, the QGD operator reduces the circuit size by deleting a single quantum gate. Smaller circuits might run faster and therefore do not timeout.

\subsubsection{Mutation score vs.\ Structural coverage criteria}

Regarding the correlation between mutation score and coverage criteria, results show a statistically significant, albeit low and positive correlation between mutation score and decision, condition, and path coverage, with Spearman's $\rho$ equal to 0.33, 0.34, and 0.33, respectively. And no correlation between mutation score and probabilistic coverage criteria.  This aligns with the classical software testing literature~\cite{CoverageVsTestEffectiveness}, which found no correlation between coverage and fault detection.

\smallskip\noindent\textbf{RQ4 (answer):} Structural coverage criteria are weakly correlated with fault detection.  While higher coverage often coincides with higher mutation scores, achieving 100\% coverage of conditions, decisions, or paths does not guarantee that faults are exposed.

\section{Discussion}\label{sec:disc}

\smallskip\noindent\textbf{Structural coverage criteria only apply to quantum controlled gates.}
Our structural coverage criteria focus on controlled quantum gates because these are the only circuit-level operations that induce conditional behavior. Uncontrolled gates always execute and therefore do not create alternative branches or path combinations; for fully sequential circuits, condition, decision, and path coverage are trivially saturated and provide limited diagnostic value. By contrast, controlled gates determine whether an operation is applied or skipped, which is the quantum analogue of classical \texttt{if}-based control flow. This focus is also practically relevant: in MQT~Bench, only 31 of 1943 circuits have no controlled gates, and in Qiskit v2.1.1, 22 of 69 built-in gate types (31\%) are controlled. Restricting our structural criteria to controlled gates, therefore, captures the main source of structural variability in real circuits while keeping the coverage definitions semantically meaningful.

\smallskip\noindent\textbf{Inherently simulation-only.}

Our current implementation is inherently simulator-centric. The \texttt{save\_\allowbreak expectation\_\allowbreak value} and \texttt{save\_\allowbreak probabilities} instructions are Qiskit Aer primitives, so \ourTool cannot compute structural coverage on real devices out-of-the-box. Moreover, \texttt{save\_\allowbreak expectation\_\allowbreak value} relies on statevector-level information, which becomes increasingly expensive as qubit counts grow.
Approximate structural coverage on hardware remains feasible, but it should be formulated as a statistical estimate rather than an exact introspection. A practical approach is to combine QMon-style instrumentation~\cite{ma_qmon_2025} (e.g., mid-circuit measurements when supported) with repeated-shot execution to estimate the activation frequencies of conditions, decisions, and paths, along with confidence intervals. This removes dependence on full statevectors and naturally incorporates noise. In this setting, coverage should be reported jointly with the shot budget and uncertainty bounds. Finally, any hardware-oriented extension must still respect the no-cloning theorem: observability should come from repeated executions and measurement design, not from cloning unknown quantum states. 

\smallskip\noindent\textbf{Coverage of the circuit vs.\ Coverage of the transpiled circuit.}
Transpilation bridges an abstract circuit and a concrete execution target, so analyzing the transpiled circuit is necessary to assess how control logic is realized at execution time. However, transpilation is backend- and configuration-dependent: different basis gates, optimization levels, or coupling maps can yield different decompositions and, consequently, different structural coverage values. Reported coverage should therefore be interpreted as \textit{target-specific structural adequacy}, not as a backend-independent property of the original high-level circuit.  

\smallskip\noindent\textbf{Weak coverage-fault detection correlation.}

RQ4's weak correlation does not make coverage uninformative; rather, it indicates that coverage and mutation score capture different qualities. Mutation score estimates fault revelation for a given mutant set, whereas structural coverage quantifies test adequacy, i.e., which controlled-gate conditions, decisions, and paths were actually exercised. In practice, coverage remains valuable as a diagnostic signal: low coverage directly exposes untested control behavior, guides targeted input generation, and supports fair comparison of test-generation strategies under the same budget. Even when two test suites achieve similar mutation scores, coverage can reveal that one explores substantially more of the circuit control structure, thereby improving robustness and reducing overfitting to a specific mutant set. Likewise, probabilistic coverage is not intended as a direct predictor of fault detection, but as an estimate of how reliably and evenly structural exploration was achieved. Coverage should therefore be interpreted as a complementary adequacy criterion rather than a substitute for fault-detection metrics.

\section{Related Work}\label{sec:rw}

Control flow in quantum circuits was formalized by Kumar~\cite{kumar_formalization_2023}, who adapted cyclomatic complexity~\cite{ciclo_complexity} to count branches in control-flow graphs of quantum circuits. These branches are static complexity abstractions, whereas our branches are execution-based and derived from controlled-gate behavior to compute condition, decision, and path coverage (including probabilistic variants).
Similarly, \citet{fortunato_gate_2024} introduced Gate Branch Coverage (GBC), which counts exercised branches of controlled gates. We preserve this intuition but revise its computation to avoid state-disturbing direct checks, reinterpret it as decision coverage, and complement it with condition and path coverage evaluated on real circuits.

\citet{wang_quito_2021, ali_assessing_2021} propose Quito, a black-box, coverage-guided approach over inputs, outputs, and input--output relations using probabilistic test oracles. Unlike Quito, our criteria target execution of internal control-flow elements (conditions, decisions, and paths) and explicitly analyze the impact of controlled gates.

\citet{ma_qmon_2025} propose QMon, which improves observability through mid-circuit measurements and defines node-, qubit-, and depth-level metrics. In contrast, we focus on control-flow-oriented coverage and controlled-gate effects. Moreover, QMon reports that monitoring coverage is constrained when preserving properties such as entanglement, whereas our simulator-based computation is not subject to the same runtime observability constraints.

\citet{shao_assessing_2026} define superposition-targeted state-space coverage for Quantum Neural Networks (QNNs), emphasizing output exploration and test-suite adequacy. Our approach instead targets structural execution coverage and extends beyond QNN-specific settings, enabling finer-grained reasoning about control flow across quantum circuits.

\citet{xia_quantum_2025} present concolic testing for Qiskit Python programs to explore classical paths and generate inputs. This is complementary to our work: they search for inputs that increase coverage, while we define quantum-specific coverage criteria for circuits for which classical path notions alone are insufficient.

\section{Conclusion and Future Work}\label{sec:concl}

This paper investigates structural coverage for circuit-based quantum programs. We show why classical criteria are insufficient in the quantum setting and introduce quantum-specific structural and probabilistic criteria for condition, decision, and path coverage.
In a large empirical study on MQT~Bench~\cite{quetschlich_mqt_2023}, we evaluated adapted classical criteria alongside our probabilistic counterparts using \ourTool. Default inputs usually achieved high condition and decision coverage but consistently under-exercised path behavior, especially in circuits with multi-controlled gates. Our criteria expose untested or rarely tested controlled-gate branches and quantify confidence in observed coverage.

As future work, we will explore the CutQC~\cite{CutQC} approach, which partitions large quantum circuits into smaller subcircuits and may enable us to extend our study to all circuits in MQT~Bench.
Furthermore, we will extend \ourTool to support additional criteria (e.g., MC/DC). We also plan to investigate input-generation tools such as QuraTest~\cite{ye_quratest_2023} and NovaQ~\cite{NovaQ}, and re-run our evaluation with different input types to assess their effect on coverage values and mutation scores.

\balance
\bibliographystyle{ACM-Reference-Format}
\bibliography{main}

\end{document}